\documentclass[apl,reprint,superscriptaddress]{revtex4-2}
\usepackage{gensymb}
\usepackage{notes2bib}
\usepackage{graphicx}
\usepackage{textcomp}
\usepackage{verbatim}
\usepackage{amsmath}
\usepackage{setspace}
\usepackage{amssymb}
\usepackage{multirow}
\usepackage[dvipsnames]{xcolor}
\usepackage[colorlinks=true,urlcolor=blue,linkcolor=blue,citecolor=blue,a4paper]{hyperref}
\graphicspath{{./images/}}

\begin{document}

\title{\large{Magnetotransport Properties of Epitaxial Films and Hall Bar Devices of the Correlated Layered Ruthenate Sr$_3$Ru$_2$O$_7$}}

\author{Prosper Ngabonziza}
\email[corresponding author: ]{pngabonziza@lsu.edu}
\affiliation{Department of Physics $\&$ Astronomy, Louisiana State University, Baton Rouge,  LA 70803, USA}
\affiliation{Department of Physics, University of Johannesburg, P.O. Box 524 Auckland Park 2006, Johannesburg, South Africa}
\author{Anand Sharma}
\affiliation{Department of Physics $\&$ Astronomy, Louisiana State University, Baton Rouge,  LA 70803, USA}
\author{Anna Scheid}
\affiliation{Max Planck Institute for Solid State Research, Heisenbergstr. 1, 70569 Stuttgart, Germany}
\author{Sethulakshmi Sajeev}
\affiliation{Department of Physics $\&$ Astronomy, Louisiana State University, Baton Rouge,  LA 70803, USA}
\author{Peter A. van Aken}
\affiliation{Max Planck Institute for Solid State Research, Heisenbergstr. 1, 70569 Stuttgart, Germany}
\author{Jochen Mannhart}
\affiliation{Max Planck Institute for Solid State Research, Heisenbergstr. 1, 70569 Stuttgart, Germany}

\date{\today}

\begin{abstract}
For epitaxial Sr$_3$Ru$_2$O$_7$ films grown by pulsed laser deposition, we report a combined structural and magnetotransport study of thin films and Hall bar devices patterned side-by-side on the same film. Structural properties of these films are investigated using X-ray diffraction and high-resolution transmission electron microscopy, and confirm that these films are epitaxially oriented and nearly phase pure. For magnetic fields applied along the $c-$axis, a positive magnetoresistance of 10\% is measured for unpatterned Sr$_3$Ru$_2$O$_7$ films, whereas for patterned Hall bar devices of channel widths of $10$ and $5\, \mu$m, magnetoresistance values of 40\% and 140\% are found, respectively. These films show switching behaviors from positive to negative magnetoresistance that are controlled by the direction of the applied magnetic field. The present results provide a promising route for achieving stable epitaxial synthesis of intermediate members of correlated layered strontium ruthenates, and for the exploration of device physics in thin films of these compounds.
\end{abstract}

\maketitle
\begin{center}
\textbf{\large{1. Introduction}}
\end{center}

Layered ruthenates are peculiar strongly correlated materials in which several comparable interactions compete to give rise to  a variety of novel electronic and magnetic phenomena. Intricate phenomena observed in layered ruthenates range from superconductivity~\cite{AChronister_2021} to emergent ferromagnetism and insulator-metal transitions~\cite{ZAli_2022}, to colossal and large magnetoresistance~\cite{EDagotto_2005}, and other electron- and spin-ordering states~\cite{SAGrigera_2001}. The layered strontium ruthenates of the  Ruddlesden-Popper (R-P) phases, Sr$_{n+1}$Ru$_n$O$_{3n+1}\, (n=1,2,3, \infty) $, play a pivotal role  in the study of strongly  correlated  electron  systems. Depending on the number $n$ of the RuO$_6$ octahedra layers in the unit cell, the phenomena obtained in these systems include potentially unconventional superconductivity in Sr$_2$RuO$_4$ $(n=1)$~\cite{AChronister_2021,YMaeno_1994,AMackenzie_2003,SAKivelson_2020}, metamagnetic quantum criticality and electron nematic fluid in Sr$_3$Ru$_2$O$_7$ $(n=2)$~\cite{RABorzi_2007,SRaghu_2009,PBMarshall_2018,PGegenwart_2006}, 
orbital-dependent double metamagnetic transition and characteristics of Hund’s metal correlations in Sr$_4$Ru$_3$O$_{10}$ $(n=3)$ (n=3)~\cite{YJJo_2006,ECarleschi_2014,DFobes_2010,PNgabonziza_2023}, and  itinerant ferromagnetism coexisting with localized correlated behavior in SrRuO$_3$  ($n=\infty$)~\cite{GKoster_2012,SHahn_2021,MKim_2015,HBoschker_2019}. The rich array of distinct collective phenomena in these materials show that Sr$_{n+1}$Ru$_n$O$_{3n+1}$ are attractive for exploring the rich physics of strongly correlated layered materials.

Beyond fundamental interests, the potential of Sr$_{n+1}$Ru$_n$O$_{3n+1}$ for applied physics has driven interests in thin films, particularly in Sr$_2$RuO$_4$ $(n=1)$ films due to its unconventional superconductivity and Shubnikov–de Haas effect~\cite{HPNair_2018,YFang_2021}, and in SrRuO$_3$  ($n=\infty$) films as an excellent electrode material integrated in diverse oxide thin-film devices for a variety of oxide electronic and nanoionic applications~\cite{GKoster_2012,PNgabonziza_2021}, and also as a candidate Weyl semimetal magnetic material~\cite{SKaneta_2022,KTakiguchi_2020}. However, due to the complexity of synthesizing phase-pure epitaxial films of intermediate Sr$_{n+1}$Ru$_n$O$_{3n+1}$ members $(1<n<\infty)$, only a few publications reported attempts to epitaxially grow  double- and triple-layered strontium ruthenate films, Sr$_3$Ru$_2$O$_7$ and Sr$_4$Ru$_3$O$_{10}$~\cite{PBMarshall_2018,WTian_2007}. Understanding the synthesis science of epitaxial films of these intermediate members provide a unique opportunity for further explorations of their correlated phases in thin films and devices. Also, the precise control of their functional interfaces at the atomic level will open new routes towards the realization of novel interface-induced functionalities in correlated layered ruthenates, which would otherwise not be accessible when using single crystals.  

We focus on epitaxial films of the bilayer strontium ruthenate Sr$_3$Ru$_2$O$_7$. The material  Sr$_3$Ru$_2$O$_7$ is  an enhanced Pauli paramagnet in  its  ground  state that undergoes an anisotropic metamagnetic transition (\textit{i.e.}, a sudden increase in the magnetization within a small change of applied magnetic field)~\cite{SIIkeda_2000,RSPerry_2001}.
Under magnetic  field  applied  in  the out-of-plane  direction, Sr$_3$Ru$_2$O$_7$ displays unusual  double metamagnetic  transitions at critical magnetic fields of $\sim 8$ T and 13.5 T~\cite{EOhmichi_2003}. However, for in-plane magnetic fields parallel to the ruthenium oxygen planes, metamagnetic behaviors shift to a lower critical field of $\sim 5.5$ T~\cite{RSPerry_2001}. In magnetotransport, a peak in the magnetoresistance (MR) has been observed around 5 T under in-plane applied magnetic fields, and it has been attributed to originate from in-plane metamagnetic behavior of Sr$_3$Ru$_2$O$_7$ ~\cite{SAGrigera_2004,WChu_2020,PBMarshall_2018}. Metamagnetic transitions in double- and triple-layered ruthenates have been proposed to arise from  magnetic fluctuations due to
the presence of flat bands near the Fermi level in the electronic band dispersions~\cite{ATamai_2008,VRShaginyan_2013,PNgabonziza_2020,PNgabonziza_2023,GGebreyesus_2022,IBenedicic_2022} 
Also, electron nematic behaviors~\cite{EFradkin_2010}, ascribed to a spin-dependent symmetry-breaking Fermi surface distortion, have been observed both in single crystals and strained epitaxial films of Sr$_3$Ru$_2$O$_7$~\cite{SAGrigera_2004,RABorzi_2007,PBMarshall_2018}. 

Although magnetransport properties of Sr$_3$Ru$_2$O$_7$ have been characterized both in bulk single crystals~\cite{JHooper_2005}, single crystalline nanosheets~\cite{WChu_2020}, and in compressively-strained epitaxial films~\cite{PBMarshall_2018}, little is known about the MR characteristics of epitaxial Sr$_3$Ru$_2$O$_7$ films and thin-film devices patterned side-by-side on the same film. Such epitaxial films are ideal for the exploration of lateral dimensional confinement effects in epitaxial films for tuning their electronic ground states. Also, having various micro- and nano-scale devices fabricated side-by-side on the same film provides an advantage over single crystal-based devices, because it provides the opportunity to perform comparative study of magnetotransport properties from the same sample, measured in similar conditions.

In  this  work,  we  report  on the epitaxy of Sr$_3$Ru$_2$O$_7$ thin films, as well as on the magnetotransport trends of epitaxial  films and Hall bar devices of Sr$_3$Ru$_2$O$_7$. The synthesis of near phase-pure Sr$_3$Ru$_2$O$_7$ films was previously reported, but only for epitaxial films grown by molecular beam epitaxy (MBE)~\cite{PBMarshall_2018,WTian_2007}. Here, we use pulsed laser deposition (PLD) for the epitaxial growth of these thin films. First, we focus on optimizing the PLD growth conditions for achieving phase-pure epitaxial Sr$_3$Ru$_2$O$_7$ films. Subsequent structural analyses of these samples, using x-ray diffraction (XRD) and high-resolution scanning transmission electron microscopy (STEM), confirm that these films are epitaxially oriented and nearly phase pure. From magnetization chacterization, the Sr$_3$Ru$_2$O$_7$ films show no ferromagnetic transition over the 
entire measured temperature range. Second, for epitaxial films prepared at optimal growth conditions, we fabricate side-by-side on the same Sr$_3$Ru$_2$O$_7$ sample several Hall bar devices of various  channel widths, to then explore their electronic transport properties. We find that the sheet resistance of Hall bar devices increases as the active channel width of Sr$_3$Ru$_2$O$_7$ films decreases, while the sheet resistance of 
unpatterned films is higher than those of structured Hall bar devices. Lastly, we perform a comparative MR study of unpatterned epitaxial Sr$_3$Ru$_2$O$_7$ films and Hall bar devices. At low temperatures, for magnetic fields applied along the $c-$axis ($B\parallel c$), with the excitation current perpendicular to the direction of applied field, MR values as high as 10\% and up to 140\% are achieved in unpatterned films and Hall bar devices, respectively. Furthermore, these films show switching behaviors from positive to negative MR that are controlled by the direction of the applied magnetic field
\begin{center}
\textbf{\large{2. Thin Film Epitaxy}}
\end{center}

\begin{figure*}[!t]
     \includegraphics[width=1\textwidth]{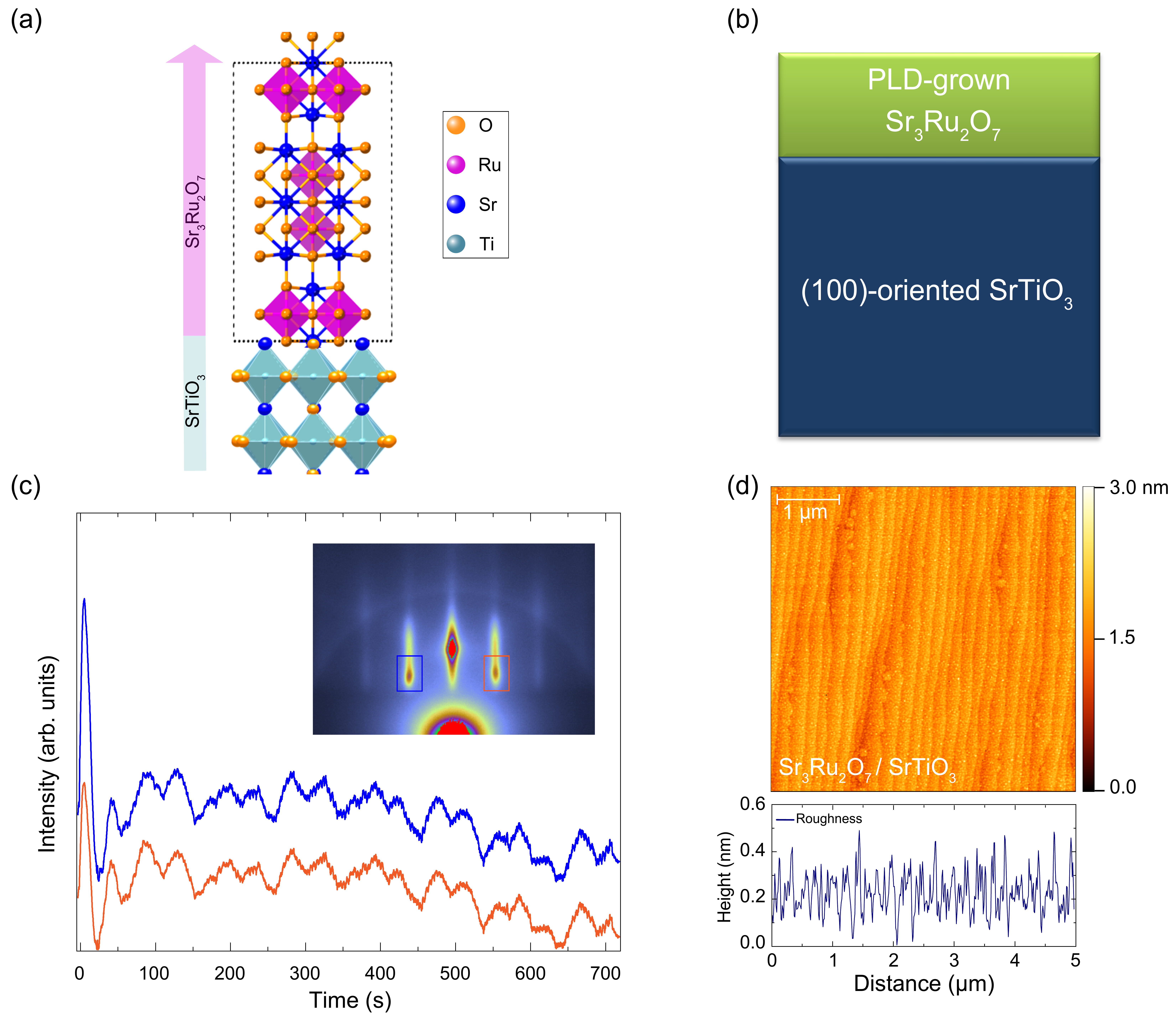}
    \caption{\textbf{Crystal structure and surface quality of epitaxial Sr$_3$Ru$_2$O$_7$ films}. Schematic  representation  of  (a) the crystal structure and (b) layout of the double-layered Sr$_3$Ru$_2$O$_7$ thin film grown by pulsed laser deposition on a (001)-oriented SrTiO$_3$ substrate. Dashed-line rectangle in (a) delineates the unit cell of Sr$_3$Ru$_2$O$_7$. (c) Time-dependent RHEED intensity oscillations recorded during the growth of 18-nm-thick Sr$_3$Ru$_2$O$_7$ film. The inset depicts RHEED patterns of a Sr$_3$Ru$_2$O$_7$ film, where the blue and orange rectangles mark the region from which the integrated intensity as a function of time was recorded during the deposition of Sr$_3$Ru$_2$O$_7$. (d) A representative AFM image displaying the surface morphology and (bottom panel) corresponding surface roughness profile of a typical Sr$_3$Ru$_2$O$_7$ film.} 
\label{Fig_01}
\end{figure*}

Figure~\ref{Fig_01}\textcolor{blue}{(a)} illustrates the epitaxy of Sr$_3$Ru$_2$O$_7$ films on a  TiO$_2$-terminated  SrTiO$_3$ substrate. All Sr$_3$Ru$_2$O$_7$ films  were grown on $(100)$-oriented SrTiO$_3$ single  crystalline  substrates  ($5 \times 5 \times 1$ mm$^3$) [Fig.~\ref{Fig_01}\textcolor{blue}{(b)}].  Prior to deposition, the SrTiO$_3$ substrates were terminated in situ using a CO$_2$ laser~\cite{WBraun_2020}. We deposited Sr$_3$Ru$_2$O$_7$ films using PLD by ablating a stoichiometric Sr$_3$Ru$_2$O$_7$ target~\cite{Note_01} with an excimer laser ($\lambda=248$ nm) at 2 Hz up to a thickness of 25 nm  at a deposition rate of $\sim 15$ \AA \, min$^{-1}$. During growth, the molecular oxygen pressure in the chamber was kept at $P_{_{\text{O}_2}}= 8.0\times 10^{-2}$ mbar. Immediately after growth, under the same $P_{_{\text{O}_2}}$, samples were cooled to room temperature at a cooling rate of 80\degree C/min. To establish stable epitaxial growth conditions, we have prepared a series of Sr$_3$Ru$_2$O$_7$ films at various substrate temperatures (T$_{\text{sub}}$), from 680 to 800\degree C in steps of 20\degree C. The optimal substrate temperature for the epitaxial growth of Sr$_3$Ru$_2$O$_7$ films was found to be T$_{\text{sub}}= 720\degree$C. At this temperature, we have also prepared a series of Sr$_3$Ru$_2$O$_7$ films varying the laser fluence from 1.5 to 3 J/cm$^2$ in steps of 0.5 J/cm$^2$, and better quality films were obtained at a laser fluence of  2.5 J/cm$^2$.  
\begin{center}
\textbf{\large{3. Results and Discussion}}
\end{center}

The surface quality of  epitaxial Sr$_3$Ru$_2$O$_7$ films was characterized by reflection  high-energy electron diffraction (RHEED) and atomic force microscopy (AFM). The  deposition  of  Sr$_3$Ru$_2$O$_7$ was in situ monitored  by  RHEED [Fig.~\ref{Fig_01}\textcolor{blue}{(c)}]. We  observe  that  the  intensity  of  time-dependent RHEED oscillations   remains  roughly the  same  throughout  the  deposition  of  the Sr$_3$Ru$_2$O$_7$ layers on SrTiO$_3$ substrates. The RHEED oscillations indicate that the Sr$_3$Ru$_2$O$_7$ films are  grown  in  a  layer-by-layer  mode  with  a  smooth  surface as  also  demonstrated  by  sharp, diffracted, and specular RHEED patterns [inset in Fig.~\ref{Fig_01}\textcolor{blue}{(c)}]. After PLD growth, the surface morphology of the Sr$_3$Ru$_2$O$_7$ films was investigated using AFM. Figure~\ref{Fig_01}\textcolor{blue}{(d)} depicts a typical AFM image for a representative 25-nm-thick Sr$_3$Ru$_2$O$_7$ film. The surface of the resulting Sr$_3$Ru$_2$O$_7$ films at optimal substrate temperature is smooth and it exhibits well-pronounced terraces, demonstrating a high quality surface morphology of these films. For a lateral scan size of $\sim 5\times 5 \, \mu \text{m}^2$, the extracted surface roughness is $\leq 0.2$ nm for a 25-nm-thick Sr$_3$Ru$_2$O$_7$ film [bottom panel in Fig.~\ref{Fig_01}\textcolor{blue}{(d)}]. 

\begin{figure*}[!t]
     \includegraphics[width=1\textwidth]{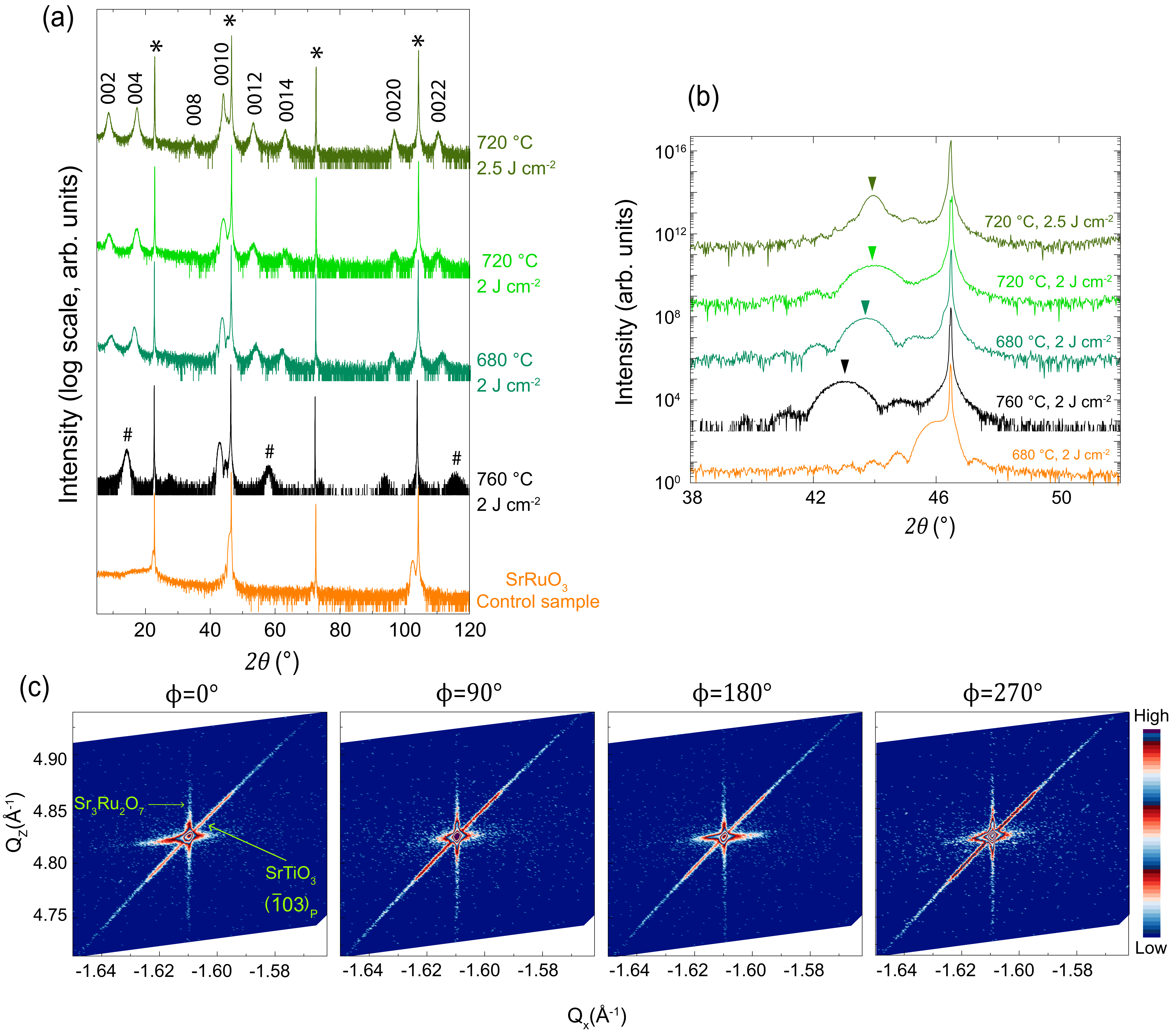}
    \caption{\textbf{Structural characterization of epitaxial Sr$_3$Ru$_2$O$_7$  films}. (a) Representative  XRD  scans of 25-nm-thick Sr$_3$Ru$_2$O$_7$ films prepared at different growth temperatures and laser fluences together with a scan of a SrRuO$_3$ control film (orange curve). For films prepared at optimal  substrate temperature ($720$\degree C), only the SrTiO$_3$ Bragg's reflection peaks (*) and the $00l$ family of the Sr$_3$Ru$_2$O$_7$ films' diffraction peaks are resolved. Extra phases ($\#$) in Sr$_3$Ru$_2$O$_7$ films are observed at high growth temperatures. (b) Closeup XRD patterns around the $00 \underline{10}$ peak of Sr$_3$Ru$_2$O$_7$. For growth temperatures higher or lower than $720$\degree C, the $00 \underline{10}$ peaks of Sr$_3$Ru$_2$O$_7$ shift to lower values. The Laue fringes in (b) indicate that there may be a thin layer of excess material, consistent with previous reports which showed that excess material in the film of the order of a few angstroms will result in Laue fringes exhibiting clear differences~\cite{AMMiller_2022}. (c) Reciprocal space maps for a Sr$_3$Ru$_2$O$_7$ film around the $\bar{1}03_{p}$ reflection of SrTiO$_3$ substrate, where $p$ refers to pseudocubic indices,  measured at four different $\phi$ angle orientations of the SrTiO$_3$ substrate.} 
\label{Fig_02}
\end{figure*}

The crystalline quality and phase purity of Sr$_3$Ru$_2$O$_7$ films were characterized by XRD. Figure~\ref{Fig_02}\textcolor{blue}{(a)}  shows representative $\theta-2\theta$ scans for the Sr$_3$Ru$_2$O$_7$ films at various substrate temperatures and laser fluence. For the films grown at optimal growth conditions (T$_{\text{sub}}= 720\degree$C and laser fluence $=2.5\, \text{J/cm}^2$), only the substrate peaks and phase-pure $00l$ family of the film diffraction peaks are resolved, which indicates a high crystallinity and verifies that the Sr$_3$Ru$_2$O$_7$ films were aligned along the c-axis. The  $002, 004, 008, 00\underline{10}, 00\underline{12}, 00\underline{14}, 00\underline{20}, \text{ and } 00\underline{22} $ Bragg diffraction peaks are well resolved, while the $006, 00\underline{16}, \text{ and } 00\underline{18} $ peaks are not present due to the small film thickness and the low structure factors of these peaks~\cite{PBMarshall_2018}. The extracted out-of-plane lattice parameter for epitaxial films grown at the optimal growth conditions is $c=20.72$ \AA . This value is consistent to the bulk  $c-$axis lattice  constant of 20.7194 \AA \, reported previously in single crystals of Sr$_3$Ru$_2$O$_7$~\cite{QHuang_1998,NAso_2005,HShaked_2000,WChu_2020}. Extra phases, indicated by $(\#)$ in Fig.~\ref{Fig_02}\textcolor{blue}{(a)}, were detected in Sr$_3$Ru$_2$O$_7$ films grown at high substrate temperatures, underlining the role of proper tuning of T$_{\text{sub}}$ for achieving phase stability in these films. These extra phases are associated with the $(00l)$ family of the diffraction peaks $(002, 008, \text{ and } 00\underline{14})$ of the $n=1$ R-P phase~\cite{MUchida_2017,JKim_2021}, in agreement with STEM data (See Fig. \textcolor{blue}{S1} in Ref.~\cite{Note-3}).

\begin{figure*}[!t]
     \includegraphics[width=0.95\textwidth]{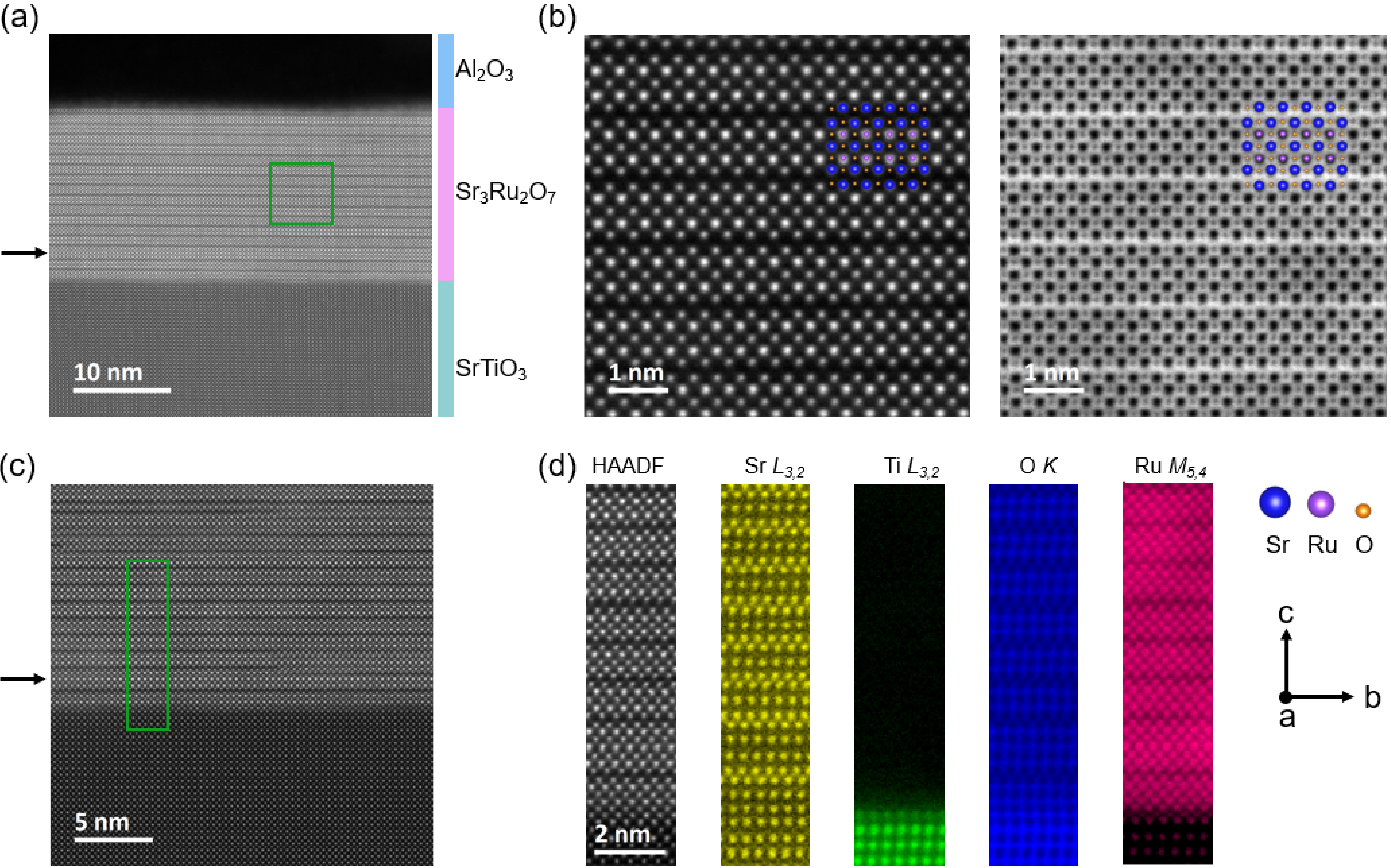}
    \caption{\textbf{Microstructural characterization of representative Sr$_3$Ru$_2$O$_7$ films prepared at  optimal  growth conditions}. (a) Overview scanning transmission electron microscopy (STEM) image of Sr$_3$Ru$_2$O$_7$ in $[100]$ zone axis orientation. (b) The magnified annular dark-field (center) and annular bright-field (right) images show the film with a good structural order. A schematic atomic structural model, displaying the sequence of Sr, Ru and O atoms is shown, overlaying the resolved double-layered atomic structure. (c) High-resolution STEM image of a Sr$_3$Ru$_2$O$_7$ film in $[100]$ orientation together with corresponding (d) electron energy-loss spectroscopy (EELS) elemental mapping of the Sr$_3$Ru$_2$O$_7$/SrTiO$_3$ interface showing the elemental distribution in the sample. Due to the overlap of Ru $M_{4,5}$-edges with low-intensity Sr $M_{2,3}$, residual noticeable Sr signals remain when extracting the Ru signal. Black arrows in (a) and (c) point to regions with $n=3$, which were observed occasionally in films prepared at optimal growth conditions} 
\label{Fig_03}
\end{figure*}

Figure~\ref{Fig_02}\textcolor{blue}{(b)} shows a closeup view of the $\theta-2\theta$ scans around the $00\underline{10}$ diffraction peak for the Sr$_3$Ru$_2$O$_7$ films grown at various T$_{\text{sub}}$ and laser fluence. For the films grown at the optimal growth conditions, only the $00\underline{10}$ peaks are resolved with noticeable thickness fringes, which highlight phase purity and smooth growth. However, we observed shifting of  the $00\underline{10}$ peaks to  lower $2\theta$ angles for films  that were not grown at the optimal growth conditions.  The  shifting  of  the  peak demonstrates  that  the  $c$-axis  lattice constant expands, indicating 
inter-growth of other Sr$_{n+1}$Ru$_n$O$_{3n+1}$ phases and change of stability of the Sr$_3$Ru$_2$O$_7$ films. In particular, the Sr$_3$Ru$_2$O$_7$ films grown at T$_{\text{sub}}=680$\degree C exhibited a peak shoulder around the substrate peak, indicating an inter-growth of higher $n$ R-P phases, with predominance of the SrRuO$_3$ phase~\cite{Note_02}. These observations are consistent with the STEM data (see, Fig.\textcolor{blue}{~S1} in Ref.~\cite{Note-3}) in which considerable inter-growths of other $n$ members of the Sr$_{n+1}$Ru$_n$O$_{3n+1}$ series were clearly detected for epitaxial Sr$_3$Ru$_2$O$_7$ films prepared outside the optimal growth conditions.

Figure~\ref{Fig_02}\textcolor{blue}{(c)} shows four reciprocal space maps (RSM) around the asymmetric pseudocubic $\bar{1}03_p$ reflection peak of the substrate for different $\phi$ angle orientations.  From all four RSMs, it is evident that the Sr$_3$Ru$_2$O$_7$ film is fully epitaxially strained and single phase, as confirmed by extracted pseudocubic in-plane lattice  constant of $a=3.889$ \AA. As the film peak is very close to the $\bar{1}03_{p}$ reflection of the substrate, potential overlap of the truncation rod from the substrate with the film peak could affect a precise extraction of in-plane lattice parameter. The extracted in-plane lattice parameter is close to reported lattice  parameter ($\sim 3.890$ \AA) of Sr$_3$Ru$_2$O$_7$ bulk single crystals~\cite{QHuang_1998,NAso_2005}.  Although the epitaxial films appear to be single phase from XRD characterization, it is also essential to use transmission electron microscopy (TEM) to check for inter-growths that are known to be difficult to discern in XRD patterns of layered R-P materials~\cite{JHHaeni_2001,EEFleck_2022,WTian_2007}.

To provide a complementary  real-space microstructural characterization of these films, cross-sectional TEM imaging was performed. With these studies, we investigated the defect populations of these films including possible inter-growths of $n\neq2$ members of the Sr$_{n+1}$Ru$_n$O$_{3n+1}$ series.  Figure~\ref{Fig_03}\textcolor{blue}{(a)} depicts a cross-sectional STEM image of the entire film thickness of a representative Sr$_3$Ru$_2$O$_7$ film prepared at the optimal growth conditions. For films prepared at the optimal substrate temperature, no surface steps at the substrate and defects throughout the entire film thickness were observed. Selected area STEM studies of these films corroborate the epitaxial orientations established by XRD and confirm that the growth occurred along the [001] direction with the $c-$axis out of plane and the films being fully epitaxially strained to the SrTiO$_3$ substrate. The STEM images of two representatives Sr$_3$Ru$_2$O$_7$ films  distinctly show the double-layered structure of the films, the interface  between  the  substrate  and  deposited  layers being coherent without misfit  dislocations  along  the  interface [Figs.~\ref{Fig_03}\textcolor{blue}{(a)} and~\ref{Fig_03}\textcolor{blue}{(c)}]. The annular dark-field (center, Fig.~\ref{Fig_03}\textcolor{blue}{(b)}) and bright-field (right, Fig.~\ref{Fig_03}\textcolor{blue}{(b)})  images  show  the  Sr and Ru columns with   enhanced   atomic   number   contrast. The Ru atomic columns appear brightest due to the much higher atomic number of Ru compared with Sr, as also indicated in the inset with a superimposed structural model [Fig.~\ref{Fig_03}\textcolor{blue}{(b)}]. The alternate stacking of the rock salt SrO layers and the RuO$_6$ octahedra sheets observed in these images confirms the formation of the desired $n=2$ member of the R-P structure. To evaluate the composition  of  the  epitaxial Sr$_3$Ru$_2$O$_7$ films,  electron  energy-loss  spectroscopy  (EELS)  measurements were performed on the cross section of the samples during STEM characterization. Figure~\ref{Fig_03}\textcolor{blue}{(d)} displays atomically resolved EELS elemental maps of a Sr$_3$Ru$_2$O$_7$ film. The black color suggests zero intensity, whereas the other colors indicate the expected chemical composition of different elements in the film (Ru,  Sr, Ti  and  O) that are resolved in the scanned region. 
\begin{figure}[!t]
     \includegraphics[width=0.48\textwidth]{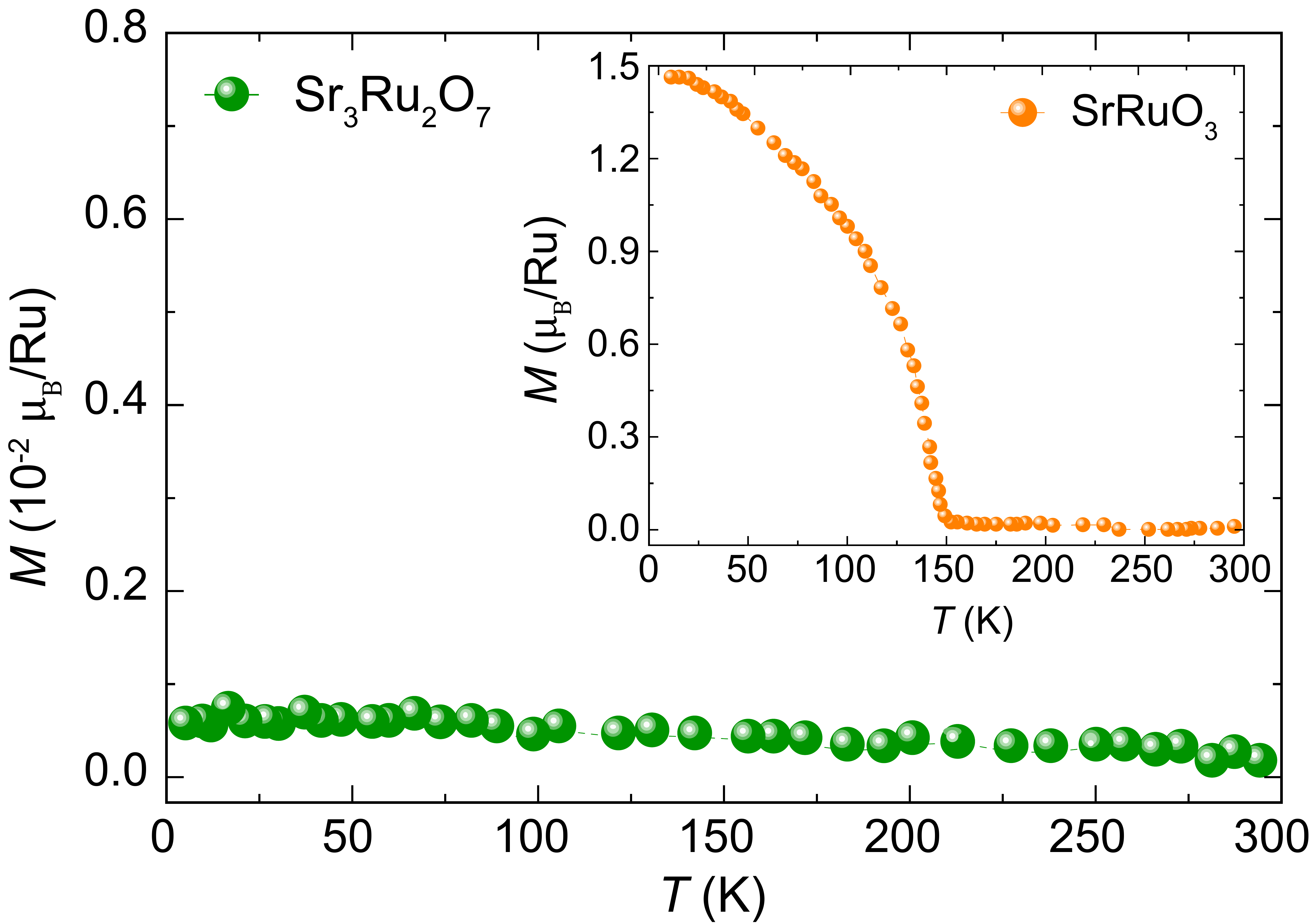}
    \caption{\textbf{Magnetic ground state of Sr$_3$Ru$_2$O$_7$ films}. Magnetization as a function of temperature of a typical Sr$_3$Ru$_2$O$_7$ film prepared at optimal growth conditions. For comparison, the magnetization data of a SrRuO$_3$ film is also presented in the inset.}
\label{Fig_04}
\end{figure}

\begin{figure*}[!t]
     \includegraphics[width=1\textwidth]{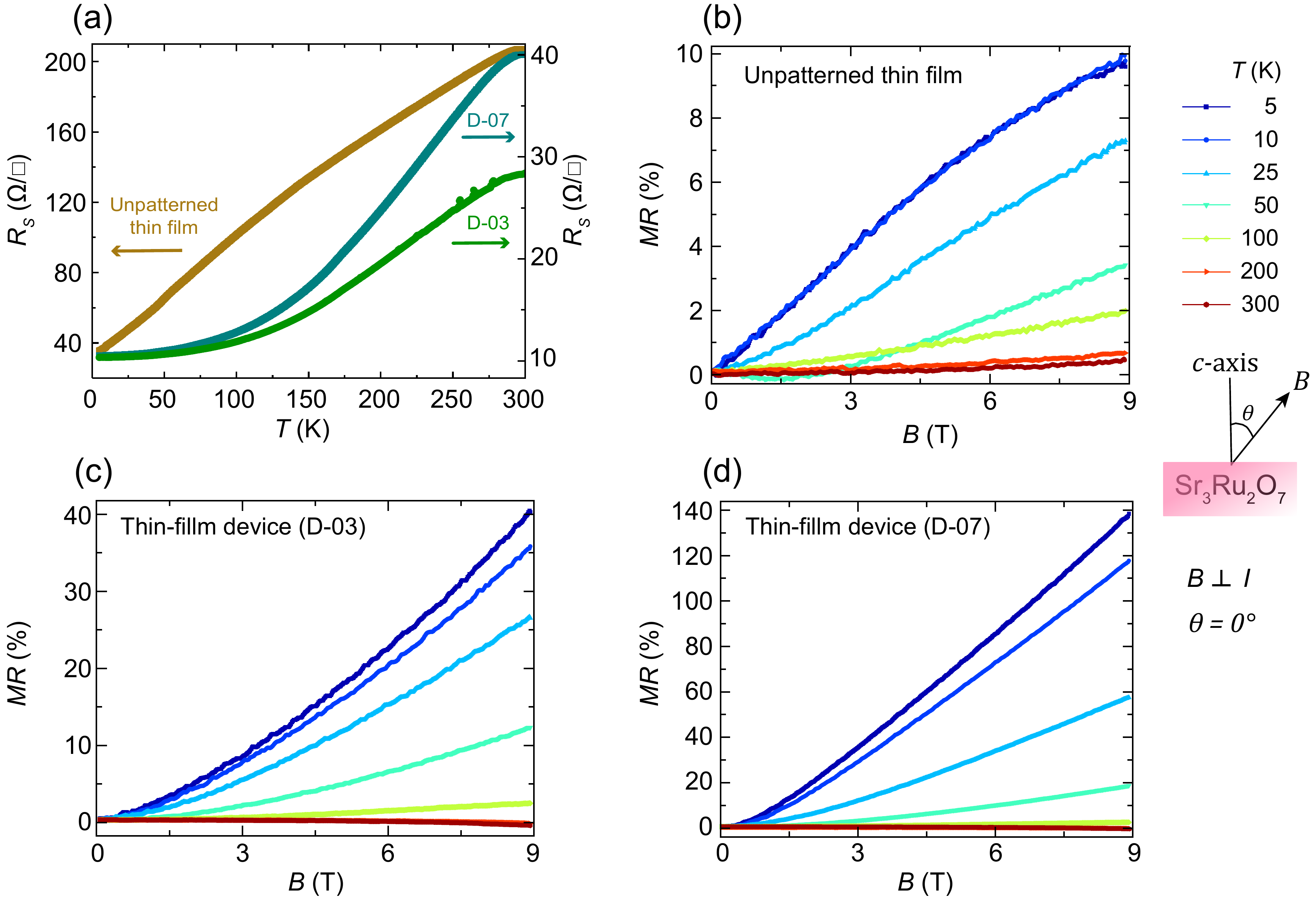}
    \caption{\textbf{Magnetotransport characteristics of a Sr$_3$Ru$_2$O$_7$ epitaxial film at various temperatures}.  (a) Temperature dependence of the zero-field sheet resistance ($R_S$) of a 25 nm-thick Sr$_3$Ru$_2$O$_7$ film before device fabrication (unpatterned thin film), and the $R_S$ of two representative patterned Hall bar devices (D-03 and D-07) fabricated on the same Sr$_3$Ru$_2$O$_7$ film. The channel widths of D-03 and D-07 are $ 10\,\mu\text{m}$ and $5\,\mu\text{m}$, respectively. (b) Magnetoresistance (MR) curves of a Sr$_3$Ru$_2$O$_7$ film measured at various temperatures in the Van der Pauw geometry before device fabrication (unpatterned thin film), and for (c)-(d) patterned Hall bar thin-film devices fabricated on the same film after measurements in the Van der Pauw configuration. MR data were acquired with $B \parallel c-$axis, $\theta = 0\degree$ ($\theta$ is  the angle between the $B-$field and $c$-axis, as schematically illustrated in the right-side of the figure). }
\label{Fig_06}
\end{figure*}

For epitaxial Sr$_3$Ru$_2$O$_7$ films grown at optimal growth conditions (T$_{\text{sub}}= 720\degree$C and laser fluence $=2.5\, \text{J/cm}^2$), the defects detected are infrequently regions of $n=3$ inter-growth within $n=2$ layers (indicated by black arrows in Figs.~\ref{Fig_03}\textcolor{blue}{(a)} and~\ref{Fig_03}\textcolor{blue}{(c)}). Such inter-growths were found in localized regions with insufficiently ordered volume to give rise to detectable diffraction spots in the XRD patterns. Similar  inter-growth of extra $n$-phases in R-P layers were reported in epitaxial films of Sr$_{n+1}$Ru$_n$O$_{3n+1}$ and Sr$_{n+1}$Ti$_n$O$_{3n+1}$ ($n=2\text{ to } 5$) grown by MBE~\cite{PBMarshall_2018,WTian_2007,JHHaeni_2001,WTian_2001}. 
STEM investigations of films grown  outside  the optimal substrate temperature window showed increased defect density and populations of 
inter-growths of other $n$-phases of the R-P series (See Fig.
\textcolor{blue}{S1} of Ref.~\cite{Note-3}). These inter-growths are difficult to control in epitaxial growth using PLD. 
This is because minor stoichiometric deviations that can exist between the target and deposited films, together with the fact that the difference in the formation energies among nearby $n$ members of the Sr$_{n+1}$Ru$_n$O$_{3n+1}$ series is small, could lead to the formation of mixed $n$-phase films of these materials~\cite{EEFleck_2022}.

Magnetization as a function of temperature, $M(\text{T})$, of Sr$_3$Ru$_2$O$_7$ films were studied using a physical property measurements systems (PPMS) vibrating sample magnetometer (VSM). The film was first pre-cooled to 5 K while in a 0.1 T, and then warmed to 300 K in the presence of 0.01 T, the $M(\text{T})$ data were collected. Figure~\ref{Fig_04} depicts the $M(\text{T})$ curve (green) of a representative Sr$_3$Ru$_2$O$_7$ film grown on SrTiO$_3$ at the optimal growth conditions. These magnetization data are obtained  after the subtraction of the diamagnetic contribution from the SrTiO$_3$ substrate. The Sr$_3$Ru$_2$O$_7$ films show no ferromagnetic transition, as the $M(\text{T})$ curve is nearly flat in the measured temperature range. This observation is in agreement with previous reports on the magnetic ground state of epitaxial films of this material~\cite{WTian_2007}. However, as it was expected, the $M(\text{T})$ data of a SrRuO$_3$ control film  prepared in the same PLD chamber show robust ferromagnetic ordering [see inset, Fig.~\ref{Fig_04}], as revealed by a sharp paramagnetic to ferromagnetic transition at round 150 K, in agreement with previous reports~\cite{WTian_2007,GKoster_2012,GLaskin_2019}. Paramagnetic behavior has also been reported in $M(\text{T})$ characteristics of high quality   Sr$_3$Ru$_2$O$_7$ crystals grown by the floating zone technique, and a notable transition from paramagnetism to ferromagnetism was observed at $\sim $ 70 K only under applied high pressure of $\sim$ 1 GPa~\cite{SIIkeda_2000}. 

Figure~\ref{Fig_06}\textcolor{blue}{(a)} presents  the  temperature-dependent  zero  magnetic field sheet resistance ($R_S$) between 5 and 300 K for a 25-nm-thick unpatterned Sr$_3$Ru$_2$O$_7$ film and for two representative patterned Hall bar devices fabricated side-by-side on this same film after Van der Pauw measurements. (For details on Hall bar devices, see Fig.~\textcolor{blue}{S3} of ~Ref.~\cite{Note-3}). The unpatterned Sr$_3$Ru$_2$O$_7$ film and devices show  metallic  behavior over the entire temperature range. At 5 K, the $R_S$  is  roughly a factor of 4 higher  in unpatterned film than in Hall bar devices. The extracted residual resistivity ratio (RRR), defined as $\rho (300\text{ K})/\rho (5 \text{ K})$, is RRR$\simeq 6.8$ for 25 nm-thick films. This value is consistent with reported resistivity data of MBE-grown Sr$_3$Ru$_2$O$_7$ films of similar thickness~\cite{PBMarshall_2018}. From the temperature dependence of the normalized resistances, $R(T)/R(5\text{ K})$, at various applied $B-$field orientations, the measured temperature dependence is isotropic in the absence of an applied field, while enhanced scattering is observed for in-plane applied $B-$field of 5 T (see, Fig.~\textcolor{blue}{S2} of ~Ref.~\cite{Note-3}). This enhanced scattering near the metamagnetic field transition was observed in a strained Sr$_3$Ru$_2$O$_7$ films~\cite{PBMarshall_2018}, and it has been attributed to field-controlled instabilities of the Fermi surface of Sr$_3$Ru$_2$O$_7$~\cite{ATamai_2008}. 

Figures~\ref{Fig_06}\textcolor{blue}{(b)}-\ref{Fig_06}\textcolor{blue}{(d)} show the MR at several temperatures for the same unpatterned film and two Hall bar devices fabricated on the same sample after measurements in Van der Pauw geometry. The magnetoresistance is  defined  as  $\text{MR}=\left(\frac{R_{\text{xx}}(B)-R_{\text{xx}}(0)}{R_{\text{xx}}(0)}\right)\times 100$, where $R_{\text{xx}} (B)$ and $R_{\text{xx}}(0)$ are the longitudinal resistance measured at a field $B$ and at zero field, respectively. MR data were acquired with the $B-$field in out-of-plane direction ($B \parallel c-$axis, $\theta=0\degree$), the excitation current being perpendicular to the field ($B \perp I$). No hysteresis loop was observed, which indicates that no intrinsic long-range ferromagnetic order occurs. Details on the analysis of MR data are discussed elsewhere~\cite{PNgabonziza_2022,PNgabonziza_2018,MPStehno_2020}

For unpatterned films measured in Van der Pauw geometry at 5 K, a positive MR is measured at low magnetic fields together with a shallow change of slope slightly above 6 T [Fig.~\ref{Fig_06}\textcolor{blue}{(b)}]. As the temperature is increased further, this shallow slope at high $B-$field disappears and MR data show quasi-quadratic field dependence. Positive MR with change of slope from a positive to a negative, resulting into a MR peak around 6 T, has been reported in thin Sr$_3$Ru$_2$O$_7$ single crystal nanosheets measured at 500 mK with $B \parallel c-$axis~\cite{WChu_2020}. The positive MR peak has been associated with the metamagnetic transition in Sr$_3$Ru$_2$O$_7$~\cite{WChu_2020,RSPerry_2001}. For patterned Hall bar devices, a positive and quasi-quadratic field dependence MR is observed independent of temperature [Fig.~\ref{Fig_06}\textcolor{blue}{(c)}-\ref{Fig_06}\textcolor{blue}{(d)}]. Unlike for bulk crystals, we are not able to resolve a MR peak in  Sr$_3$Ru$_2$O$_7$ thin films due most probably to impurity in these films, and also because MR data were acquired only down to 5 K and up to 9 T. For Sr$_3$Ru$_2$O$_7$ single crystals, a broad MR peak at 5 K was resolved only for MR data acquired up to 15 T~\cite{RSPerry_2001}. 

The MR increases with decreasing temperature and it evolves into quasi-linear
behavior in the high field limit. For the unpatterned Sr$_3$Ru$_2$O$_7$ film, a maximum MR $(T=5 \text{ K}, B= 9\text{ T})$ of 10\% is obtained [Fig.~\ref{Fig_06}\textcolor{blue}{(b)}]. However, Hall bar devices fabricated on this same film show large positive MR  values of 40\% and 140\% for devices of channel widths of $10\text{ and } 5\, \mu\text{m}$, respectively [Fig.~\ref{Fig_06}\textcolor{blue}{(c)}-\ref{Fig_06}\textcolor{blue}{(d)}]. The positive MR for $B\parallel c$ is an indication of paramagnetic behavior of these films, which is consistent with known resistivity increase as magnetic field
increases in most materials~\cite{RNiu_2022}. While for ferromagnets, MR is negative; for paramagnetic metals, the movement of carriers will be deflected under a Lorentz force, which increases the probability of carrier scattering, thereby increasing the resistance. 

The differences revealed by the transport measurement results [Fig.~\ref{Fig_06}\textcolor{blue}{(a)}] and MR values of unpatterned films and patterned Hall bars [Fig.~\ref{Fig_06}\textcolor{blue}{(b)}-\textcolor{blue}{(d)}] is plausibly caused by  inhomogeneities in the electron concentration caused by disorder in the sample. This will cause the bulk region to present different types of local current density and different potential landscape, resulting in different evolutions of the longitudinal potential profiles over certain portion of the film. We note that such types of inhomogeneities  in  disordered  conductors is known to give rise  to  large MR,  as  described  by  the  Parish  and Littlewood  model~\cite{MMParish_2003,MMParish_2005} that was used  to  explain  large MR values in a range of materials (\textit{e.g.},  Ag$_{2\pm\delta}$Te, Ag$_{2\pm\delta}$Se, and Bi$_2$Te$_3$). In this case, the materials were  modeled  as  a  network  of  resistors  due  to disorder-induced mobility fluctuations~\cite{MMParish_2003,DLeusink_2014,deBoer_2019}. For Hall bars of various channel widths structured on a Sr$_3$Ru$_2$O$_7$ film with some inhomogeneous disordered conduction (\textit{e.g.}, regions with higher conductivity than others), there will be inhomogeneity in the electron concentration over a small area in Hall bars, thus  resulting in larger MR values in Hall bar devices as compared to unpatterned Sr$_3$Ru$_2$O$_7$ films. This interpretation is consistent to the observation that the Hall bar device with narrow channel width (D-07 of 5 µm width) exhibit large resistance and MR values as compared to the Hall bar device D-03 that has a channel width of 10 µm.

\begin{figure}[!t]
     \includegraphics[width=0.48\textwidth]{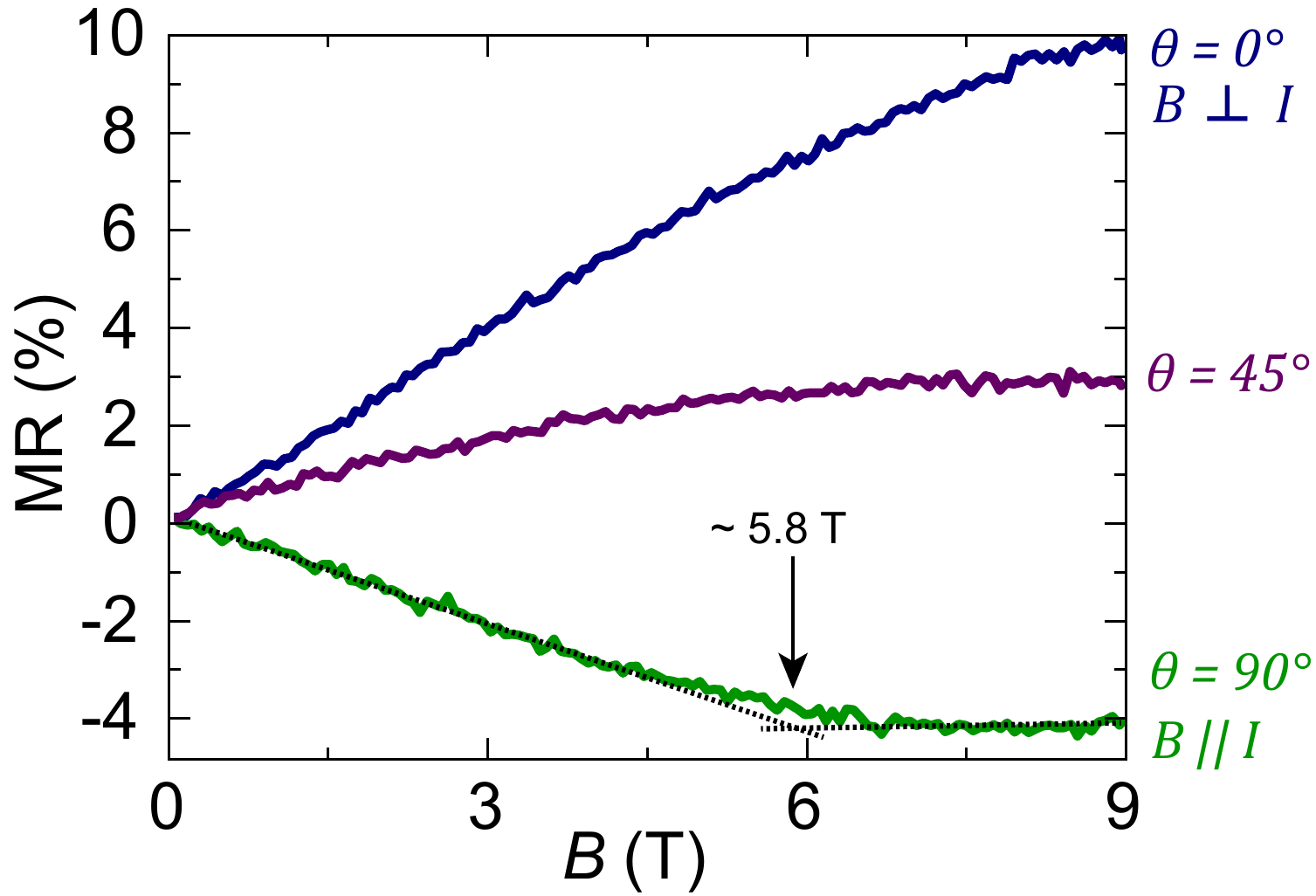}
    \caption{\textbf{MR characteristics at different $B-$field orientation}. The MR data of unpapatterned  Sr$_3$Ru$_2$O$_7$ film were measured at 5 K in the Van der Pauw geometry. Linear fits (black dotted lines) on the MR curve for $\theta =90\degree$ indicate a change of slope at a magnetic field of $\sim 5.8$ T.}
\label{Fig_07}
\end{figure}

For the same film characterized in Figure~\ref{Fig_06}, we have performed angle dependence MR measurements at various temperatures. Figure~\ref{Fig_07} depicts representative MR data at 5 K for unpatterned Sr$_3$Ru$_2$O$_7$ film. The MR data were acquired for $\theta$ (the angle between the $B-$field and the $c-$axis of the sample) values of $0\degree, 45\degree \text{ and } 90\degree$. As $\theta$ increases from $0\degree \text{ to } 90\degree$, the MR undergoes a gradual transition from positive to negative. For $\theta=90$, the MR is negative and it reaches a local minimum at a magnetic field of $\simeq 5.8$  T, which is consistent with the in-plane metamagnetic transition behavior~\cite{PBMarshall_2018}. MR data of Hall bar device fabricated on the same film exhibit qualitatively 
similar switching behaviors from positive to negative MR as the magnetic field is applied from out-of-plane ($B\parallel c-$axis) to in-plane direction ($B \parallel ab-$plane) (see Fig.~\textcolor{blue}{S4} of Ref.\cite{Note-3}). A maximum negative MR $(T=5 \text{ K}, B= 9\text{ T},\theta=90\degree)$ values of -4\% and -70\% are obtained for unpatterned film [Fig.~\ref{Fig_07}] and Hall bar device (see Fig.\textcolor{blue}{~S4} of Ref.\cite{Note-3}), respectively. 

The observation of a negative MR at $\theta= 90\degree$, accompanied with a local minimum and shallow positive slope at high magnetic field, is attributed to the coexistence of short-range in-plane ferromagnetic order in the surface layer and metamagnetism in these films. The Ru$^{4+}$ spin becomes aligned due to the applied magnetic field, thus spin-dependent scattering decreases, which gives rise to a negative MR. This observation is consistent with a recent magnetic field-controlled spectroscopic study that resolved in-plane ferromagnetism in Sr$_3$Ru$_2$O$_7$ single crystals that is sensitive to the direction of applied magnetic field~\cite{MNaritsuka_2023}. Furthermore, we conjecture  that 
the high MR values in these films could be due to spin fluctuations of the mobile electronic carriers when the material is under applied magnetic fields.

While nematic phases have been been reported in bulk single crystals of Sr$_3$Ru$_2$O$_7$ at low temperatures ($< 1$ K)~\cite{RABorzi_2007,JANBruin_2013,CStingl_2011}, there are no clear signatures of nematic phases detected in magnetotranpsort data of these PLD-grown films. The absence of nematic phase features in these epitaxial films is attributed to enhanced defect density, as evidenced by extracted RRR and TEM data. Further improvements of the crystalline quality of Sr$_3$Ru$_2$O$_7$ films, by reducing impurities and carrier scattering effects, will lead to an increase in RRR towards the bulk corresponding value of $\sim 80$~\cite{WChu_2020,SIIkeda_2000}, as well as to the detection of clear signatures of novel emergent phenomena (\textit{e.g.}, nematic phase and non-Fermi-liquid behavior) in low-temperature magnetotransport data of Sr$_3$Ru$_2$O$_7$ films and patterned Hall bar devices.

In summary, we have explored an approach to stabilize the epitaxy  of Sr$_3$Ru$_2$O$_7$ films by varying the substrate temperature and laser fluence in the PLD process, and investigated the magnetotransport trends in these epitaxial films. Structural characterizations of Sr$_3$Ru$_2$O$_7$ films prepared at optimal growth conditions confirm that these films are epitaxially oriented and nearly phase pure. Resistivity measurements of Sr$_3$Ru$_2$O$_7$ films and Hall bar devices exhibit characteristic metallic behavior. The magnetization characteristics of  the films show no intrinsic ferromagnetic transition over the entire measured temperature range. These films exhibit a qualitatively similar MR behavior as previously reported for bulk single crystals and strained epitaxial films, specifically positive MR for the magnetic field applied along the $c-$axis. However, we achieve unprecedented large magnetoresistance values up to 140\% in Hall bar devices patterned on Sr$_3$Ru$_2$O$_7$ epitaxial films. 
Briefly, this work is not only the first reported effort for the epitaxial growth of nearly phase-pure Sr$_3$Ru$_2$O$_7$ films by PLD, but also, it is the first endeavor for the exploration of magnetotransport characteristics of a series of Hall bar devices patterned side-by-side on the same film, which help define a pathway towards understanding the synthesis science and device physics of epitaxial Sr$_3$Ru$_2$O$_7$ films. This provides an opportunity to explore more the physics of Sr$_3$Ru$_2$O$_7$-based devices for the realization and modulation of novel emergent phenomena at low temperatures in phase pure films. Based on these results, future directions are expected to focus on improving further the films quality and the investigation of magnetotransport properties of quantum structures ($e.g.,$ nonowires and quantum dots) patterned on epitaxial Sr$_{n+1}$Ru$_n$O$_{3n+1}$ films, as well as on nanoscale devices  fabricated side-by-side on the same film in which magnetotransport properties can be modulated through quantum size effects and applied electric fields.

\textbf{\large{Acknowledgments}}

The  authors  acknowledge technical support from Sarah Parks.

P. Ngabonziza acknowledges startup funding from the College of Science and the Department of Physics \& Astronomy at Louisiana State University. 

A. Scheid acknowledges the invaluable assistance of Tobias Heil for his support with STEM examinations, and Y. Eren Suyolcu for his exceptional efforts in coordinating the STEM examinations and interpreting the results.

A. Scheid and P. van Aken acknowledge funding from the European  Union’s  Horizon  2020  research  and  innovation programme under Grant Agreement No.823717-ESTEEM3.
\newline \newline
\textbf{\large{Data Availability}}

The data that support the findings of this study are available from the corresponding author upon reasonable request.

\bibliography{references_SRO-327}

\onecolumngrid
\newpage
\setcounter{table}{0}
\setcounter{figure}{0}
\renewcommand{\thefigure}{S\arabic{figure}}%
\setcounter{equation}{0}
\renewcommand{\theequation}{S\arabic{equation}}%
\setstretch{1.25}
\begin{center}
\title*{\textbf{\large{Supplementary Information:} \\ [0.25in] \large{Magnetotransport Properties of Epitaxial Films and Hall Bar Devices of the Correlated Layered Ruthenate Sr$_3$Ru$_2$O$_7$}}}
\end{center}
\begin{center}
\large{Prosper Ngabonziza,$^{1,2,{\textcolor{blue}{\small{*}}}}$ Anand Sharma,$^1$  Anna Scheid,$^3$ Sethulakshmi Sajeev,$^1$ \\ \; \;\;\;\;\;\;\;\;\;\;\;\;\;\;\;Peter A. van Aken,$^3$ and Jochen Mannhart$^3$}\newline 
\\
\large{\textit{$^1$Department of Physics $\&$ Astronomy , Louisiana State University, Baton Rouge, LA 70803, USA
\newline $^2$Department of Physics, University of Johannesburg, P.O. Box 524,  Auckland Park 2006, \\ \; \;\; \;\; \;\; \;\; \;\; \;\; \;\; \;\; \;\; \;\;Johannesburg, South Africa \newline $^3$Max Planck Institute for Solid State Research, Heisenbergstr. 1, 70569 Stuttgart, Germany \newline}}
\end{center}

\begin{center}
\textbf{\large{Microstructural  Characterization of  Epitaxial Sr$_3$Ru$_2$O$_7$ Films}}
\end{center}

Scanning transmission electron microscopy (STEM) and electron energy loss spectroscopy
(EELS) investigations were performed using a Cs-probe-corrected JEOL JEM-ARM200F.\begin{figure*}[!h]
     \includegraphics[width=0.825\textwidth]{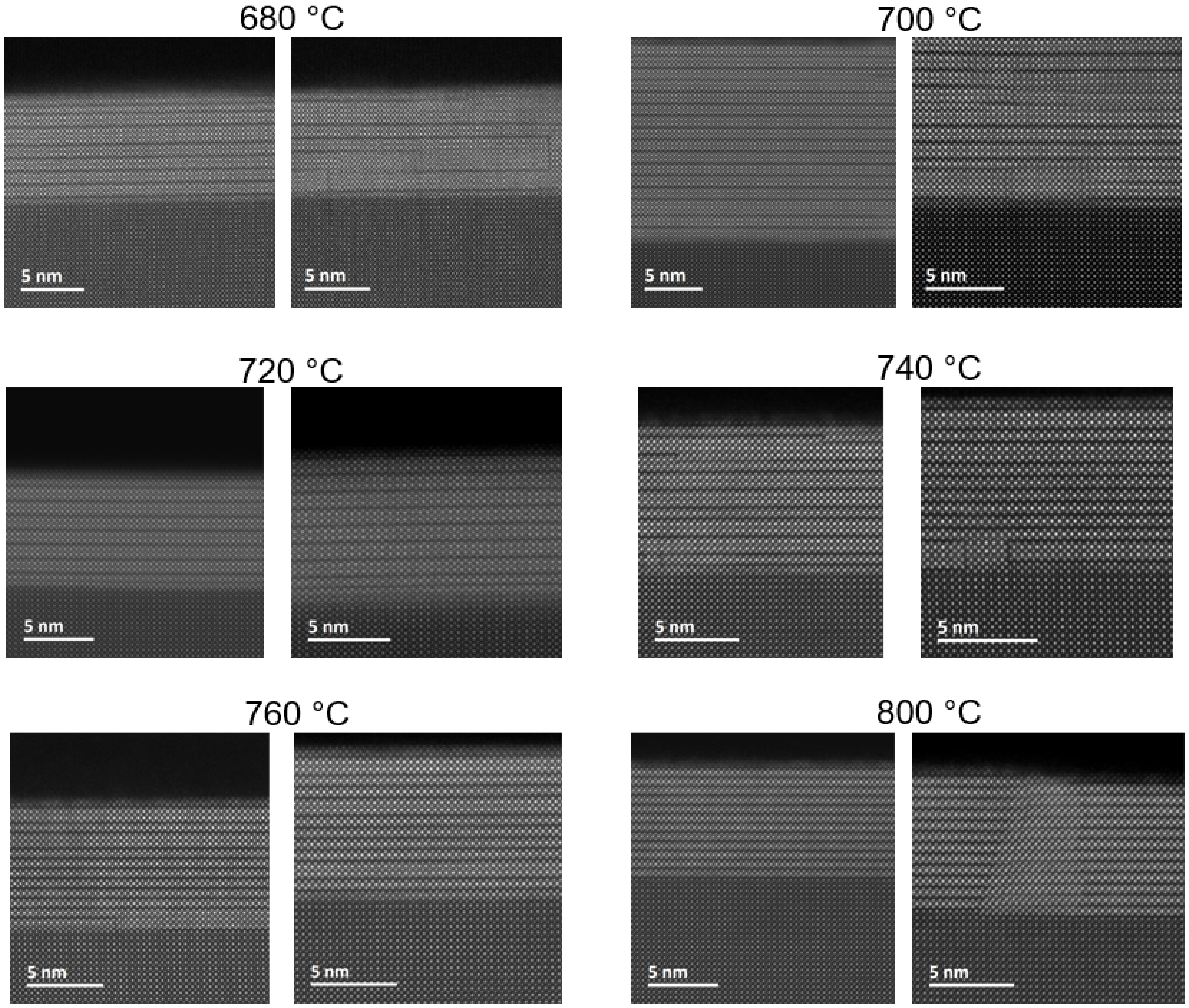}
    \caption{STEM images of Sr$_3$Ru$_2$O$_7$ films grown on (100)-oriented SrTiO$_3$ substrates at  various substrate temperatures (T$_{\text{sub}}$). }
\label{Fig_S1}
\end{figure*} Using STEM, we have performed a detailed microstructural characterization of Sr$_3$Ru$_2$O$_7$ films grown by pulsed laser deposition (PLD) at different substrate temperature. To improve the signal-to-noise ratio of the Electron energy-loss spectroscopy (EELS) elemental mapping of Sr$_3$Ru$_2$O$_7$/SrTiO$_3$ interface [see, Fig.~\textcolor{blue}{3(d)} of the main text], we used principal component analysis including 20 principal components. Figure~\ref{Fig_S1} depicts representative STEM images in $[100]$ zone axis orientation for Sr$_3$Ru$_2$O$_7$ films grown at various substrate temperatures ranging from 680\degree C to 800\degree C. At a substrate temperature of 680\degree C, the films exhibit some intergrowth and noticeable defects. As the growth temperature increases,  intergrowth and structural defects reduces and at 720\degree C, films exhibit a dominant  Sr$_3$Ru$_2$O$_7$ phase with no noticeable defects in the scanned areas. The optimal substrate temperature of Sr$_3$Ru$_2$O$_7$ films grown by PLD on SrTiO$_3$  substrates was found to be around 720\degree C. Increasing further the substrate temperature above 720\degree C, the films exhibit increased defect density and considerable inter-growths of other $n$ members of the Sr$_{n+1}$Ru$_n$O$_{3n+1}$ Ruddlesden-Popper (RP) phases are observed. In general we observe a trend that for lower growth temperatures, the  $n\geq 2$ RP phases predominate in the parasitic phases, whereby this fraction decreases with increasing temperature until finally at higher substrate temperatures (800\degree C), the films are dominantly $n=1$ RP phase and in the scanned areas they exhibit defects running across the film from the interface between substrate and the film.
\begin{figure*}[!t]
     \includegraphics[width=0.9\textwidth]{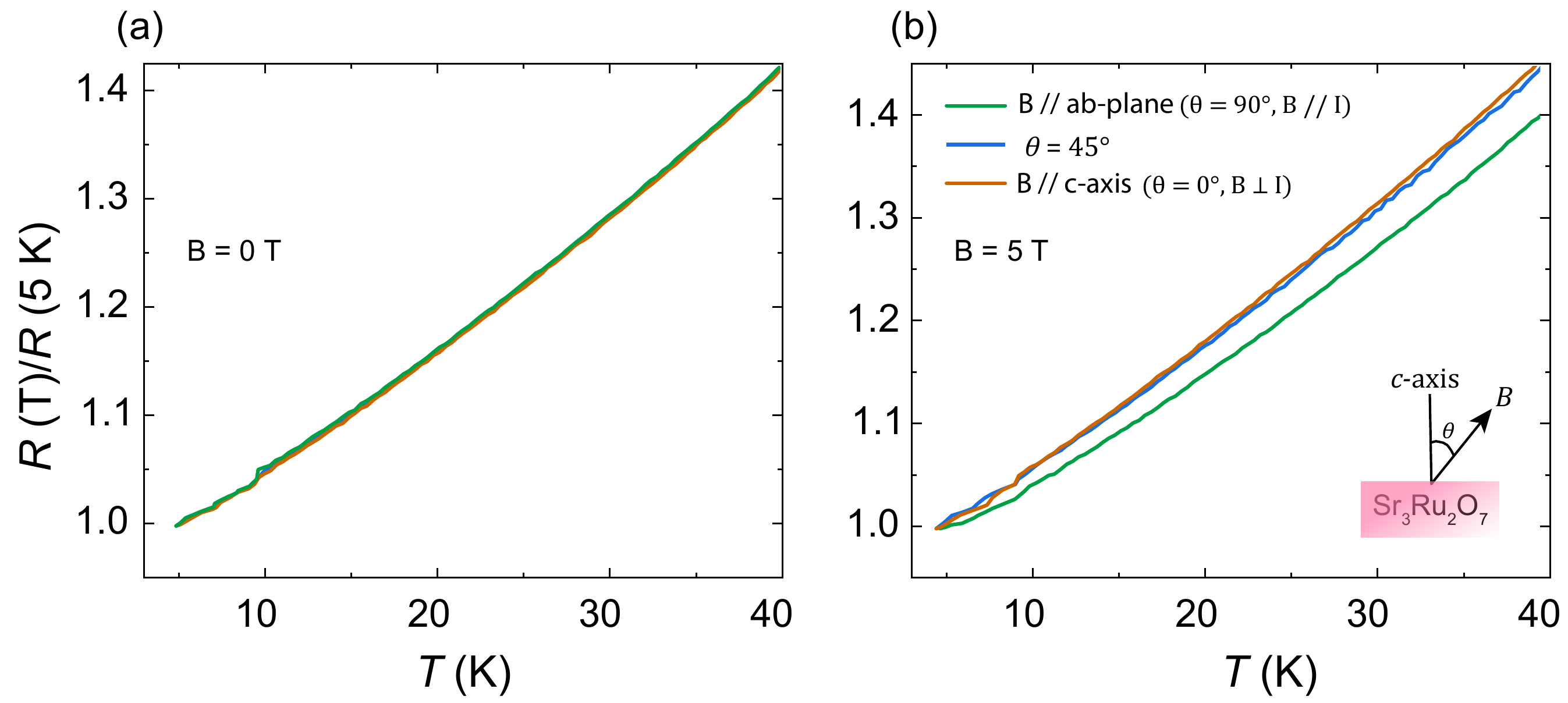}
    \caption{Normalized sheet resistance of epitaxial Sr$_3$Ru$_2$O$_7$ films with an applied magnetic field of (a) 0 T and (b) 5 T. The applied field $B$ is along various directions: $B-$field parallel to the $c-$axis ($\theta = 0\degree$) with the current perpendicular to the field ($B \perp I$), $B-$field parallel to the $ab-$plane ($\theta = 90\degree$) with the current parallel to the field ($B \parallel I$), and in the configuration of the $B-$field applied at an angle of $\theta=$45\degree \,with respect to the $c-$axis.}
\label{Fig_S2}
\end{figure*}

For device fabrication and electronic transport characteristics of epitaxial Sr$_3$Ru$_2$O$_7$ films prepared at the optimal growth conditions, we fabricated side-by-side on the same film a series of Hall bar devices of various channel widths (from 10 $\mu\text{m}$ to 500 nm) [Fig. \textcolor{blue}{S3(a)}]. The devices were patterned using electron-beam lithography (EBL) followed by ion-beam etching and the deposition of Ti/Au (5 nm/45 nm) electrodes. Figures \textcolor{blue}{S3(b)} and \textcolor{blue}{S3(c)} show top-view scanning electron microscopy (SEM) images of patterned Hall bar devices and a schematic side view of a device together with transport measurement configuration, respectively. Unpatterned Sr$_3$Ru$_2$O$_7$ films were measured in a Van der
Pauw configuration by wire bonding aluminum wires to the samples’ corners. Electronic transport characteristics of
patterned thin-film devices, which were fabricated side-by-side on the same film after transport measurements in a
Van der Pauw geometry, were studied in a Hall bar configuration with aluminum wires connected to the electrodes
[Fig. \textcolor{blue}{S3(d)}]. For magnetotransport measurements, we used a Quantum Design PPMS operated at 300-5 K. An
excitation current of 1 $\mu$A was applied. The MR measurements were performed with the magnetic field (B) oriented-both along the c-axis (B $\parallel$ c) and in the ab-plane (B$\parallel$ ab) of the film.
\begin{figure*}[t]
     \includegraphics[width=0.95\textwidth]{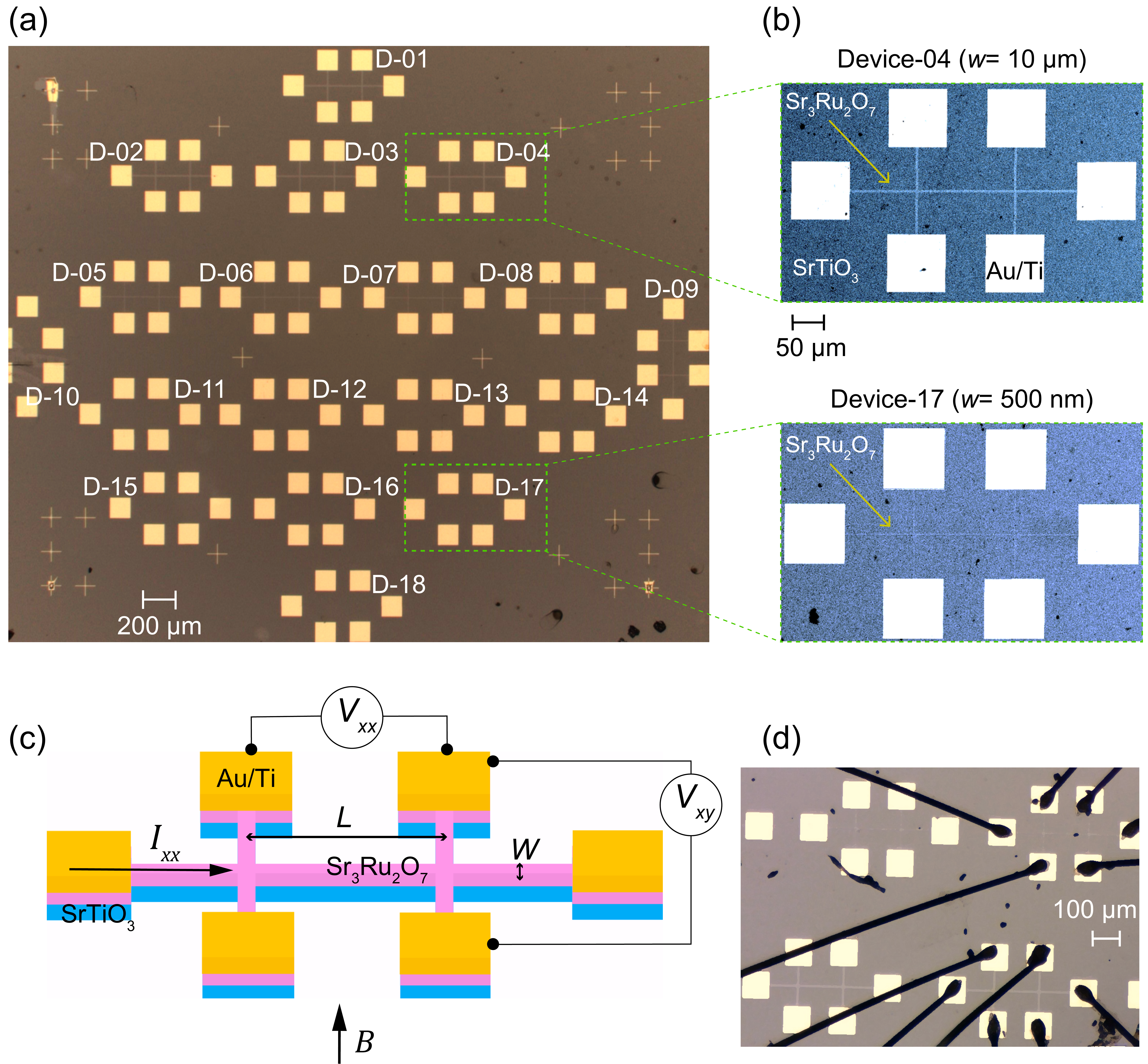}
    \caption{\textbf{Device fabrication and measurement configuration}. (a) Optical micrograph image showing a top view of the whole Sr$_3$Ru$_2$O$_7$ film on which several Hall bar devices of different channel widths $(W)$ were fabricated side-by-side on the same film. The channel widths are $W=$ $ 10\,\mu\text{m}$, $5\,\mu\text{m}$, $1\,\mu\text{m}$ and $500$ nm for Hall bar devices D-01 to D-04, D-05 to D-09, D-10 to D-14, and D-15 to D-18, respectively. (b) SEM images of representative Hall bar devices of $W = 10\,\mu\text{m}$  and 500 nm. (c) Configuration of magnetotransport measurement for Sr$_3$Ru$_2$O$_7$ Hall bar devices. The length ($L$) between longitudinal voltage contacts is the same for all Hall bar devices in (a), $L=200\,\mu\text{m}$. (d) Photograph of Sr$_3$Ru$_2$O$_7$ devices that are wire-bonded for magnetotransport measurements. Due to the limited number of available contacts on a chip carrier, only two devices on a sample could be bonded at once.} 
\label{Fig_05}
\end{figure*}

\begin{figure*}[!h]
\centering
\begin{minipage}[c]{.5\linewidth} 
\includegraphics[width=1\textwidth]{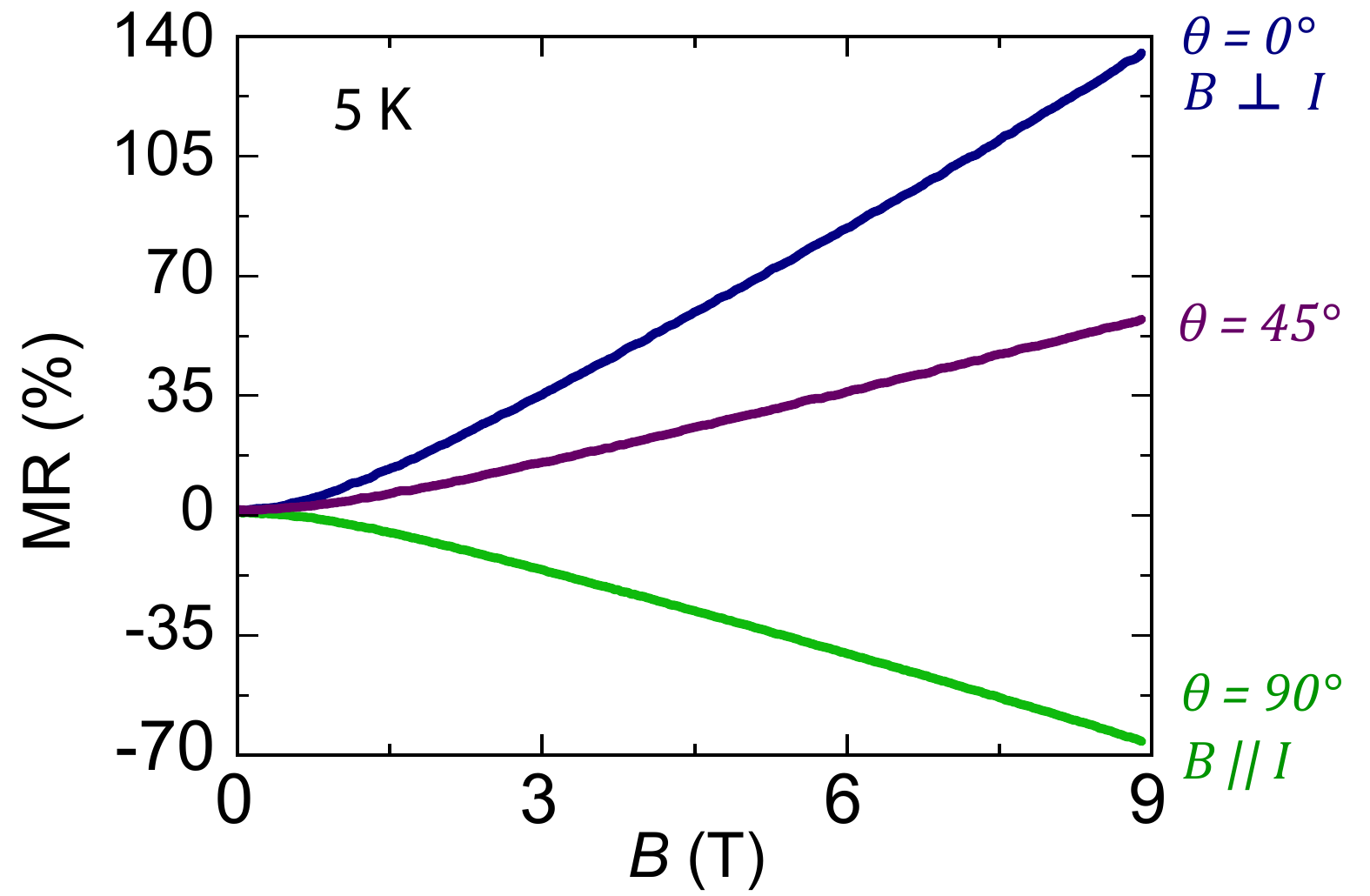}
\end{minipage}
\begin{minipage}[c]{.25\linewidth} 
\caption{\linespread{1.25} The MR data at different orientation of applied magnetic field of a patterned Hall bar device fabricated on the same film presented in Fig.~\textcolor{blue}{6} of the main text. Device fabrication was done after measurements in the Van der Pauw configuration.}
\label{Fig_S3}
\end{minipage}
\end{figure*}

Figure~\ref{Fig_S2} depicts the temperature dependence of the normalized sheet resistance, $R(\text{T})/R(\text{5 K})$, of a representative 25-nm-thick Sr$_3$Ru$_2$O$_7$ film acquired with applied magnetic field of 0 T [Fig.~\ref{Fig_S2}\textcolor{blue}{(a)}] and 5 T [Fig.~\ref{Fig_S2}\textcolor{blue}{(b)}]. Data were taken with the magnetic field applied along the the $c-$axis ($B \parallel c-$axis), along the $ab-$plane ($B \parallel ab-$plane), and the field at an angle of 45\degree \, with respect to the $c-$axis. In the absence of an applied magnetic field, the normalized sheet resistance show similar characteristics with the curves nearly overlapping on top of one another [Fig.~\ref{Fig_S2}\textcolor{blue}{(a)}]. At the critical magnetic field of 5 T, enhanced scattering is observed when the excitation current is parallel to applied magnetic field [Fig.~\ref{Fig_S2}\textcolor{blue}{(b)}]. 
Figure~\ref{Fig_S3} depicts MR data at 5 K for a patterned Sr$_3$Ru$_2$O$_7$ Hall bar device (D-07 of channel width of $5\mu$m). The MR data were acquired for $\theta$ (the angle between the $B-$field and $c$-axis, as schematically illustrated in the inset of Fig.~\ref{Fig_S2}) values of $\theta = 0\degree, 45\degree, \text{ and } 90\degree$.

\end{document}